\documentclass[pra,twocolumn]{revtex4}
\usepackage{amsmath,amsfonts,amssymb}
\usepackage{graphicx,float,calc}
\usepackage{color,bm}
\usepackage{ulem}
\usepackage[colorlinks,urlcolor=blue,citecolor=blue,linkcolor=blue]{hyperref}

\begin{document}
\author{Zhen Zheng$^{1}$}
\author{Han Pu$^{2,3}$}
\thanks{hpu@rice.edu}
\author{Xubo Zou$^{1}$}
\thanks{xbz@ustc.edu.cn}
\author{Guangcan Guo$^{1}$}
\affiliation{$^1$Key Laboratory of Quantum Information, and Synergetic Innovation Center of Quantum Information \& Quantum Physics, University of Science and Technology of China, Hefei, Anhui, 230026, People's Republic of China}
\affiliation{$^2$Department of Physics and Astronomy, and Rice Center for Quantum Materials, Rice University, Houston, TX 77251-1892, USA}
\affiliation{$^3$Center for Cold Atom Physics, Chinese Academy of Sciences, Wuhan 430071, China}
\title{Artificial topological models based on a one-dimensional spin-dependent optical lattice}

\begin{abstract}

Topological matter is a popular topic in both condensed matter and cold atom research.
In the past decades, a variety of models have been identified with fascinating topological features. Some, but not all, of the models can be found in materials.
As a fully controllable system, cold atoms trapped in optical lattices provide
an ideal platform to simulate and realize these topological models.
Here we present a proposal for synthesizing topological models in cold atoms
based on a one-dimensional (1D) spin-dependent optical lattice potential. In our system, features such as staggered tunneling, staggered Zeeman field, nearest-neighbor interaction, beyond-near-neighbor tunneling, etc. can be readily realized.
They underlie the emergence of various topological phases.
Our proposal can be realized with current technology and hence has potential applications
in quantum simulation of topological matter.

\end{abstract}
\maketitle

\section{Introduction}

Topological materials have attracted intensive attention in recent years.
They are characterized by topological protected edge states,
which emerge as gapless excitations localized at material boundaries.
A series of investigations on topological materials have shed new light on the research of
a new classification paradigm based on the topological order \cite{topo-cri-1,topo-cri-2,topo-cri-3},
and promises potential applications such as topological quantum computation \cite{intro-1}. However, many interesting topological models cannot be readily found in natural materials.
As a result, the realization of topological models in a well-controlled system is highly desirable.

Compared with conventional solid-state systems,
cold atoms offer a perfectly clean platform with great controllability to study topological models.
It possesses the following important advantages:
(i) Feshbach resonance makes it possible to manipulate atomic interactions by external magnetic or optical fields \cite{feshbach}.
(ii) Laser fields that couple hyperfine states of atoms can synthesize effective physical fields,
such as Zeeman fields, and spin-orbital couplings \cite{gauge-field}.
The strengths of those synthetic fields are determined by the laser beams and hence are tunable.
(iii) The configuration of an optical lattice can be designed via several counter-propagating lasers.
For example, the current technique has realized spin-dependent lattice systems \cite{sp-lattice-1,sp-lattice-2}
and various unconventional lattice potentials \cite{exotic-lattice-1,exotic-lattice-2}.

In this paper, we propose a scheme for synthesizing topological models in cold atoms.
Following our earlier work \cite{p-wave}, the key idea in this proposal
relies on a spin-dependent optical lattice potential consisting of two sublattices with spatial offset.
Our proposal can exhibit the topological transition via tunable experimental parameters,
and thus provides a versatile platform to study their intrinsic topological properties. In our previous work \cite{p-wave}, we focused on realizing nearest-neighbor effective $p$-wave interaction in a system of $s$-wave interacting fermions. In the present work, we focus instead on manipulating single-particle physics in such a system.

The paper is organized as follows.
In Sec.~\ref{sec-general-model}, we describe the general spin-dependent lattice model,
and show its realization via current experimental techniques.
Sec.~\ref{sec-topo} shows the topologically nontrivial properties of our lattice potential,
which demonstrates our lattice potential as a promising candidate to realize a wide variety of topological models.
The measurement of the topological invariant is discussed in Sec.~\ref{sec-measure}.
In Sec.~\ref{sec-site}, we study the edge states associated with the topological phase,
and give an explanation to the boundary effect.
In Sec.~\ref{sec-flux}, we extend the idea of our proposal to synthesize artificial magnetic field
using the laser-assisted tunneling technique.
In Sec.~\ref{sec-summary}, the extension in the presence of interactions is discussed.

\section{1D lattice model} \label{sec-general-model}

\begin{figure}
\centering
\includegraphics[width=0.48\textwidth]{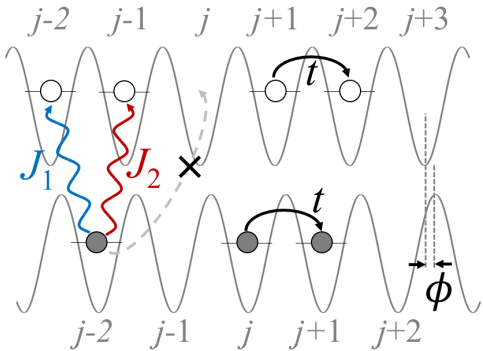}
\caption{(Color online)
Optical lattice configuration: the empty and solid circles represent the spin $\uparrow$ and $\downarrow$ atoms, respectively.
The black solid lines describe the conventional tunneling with the amplitude $t$.
The blue and red solid wavy lines describe the rf-induced tunneling with the amplitude $J_1$ and $J_2$, respectively.
The inter-sublattice tunneling marked by the gray dashed line is considered to be negligible.}
\label{fig-lattice}
\end{figure}

We start from a degenerate Fermi gas with two hyperfine states denoted as \textit{pseudospins} ($\uparrow$ and $\downarrow$) trapped in a one-dimensional (1D) optical lattice as illustrated in Fig.~\ref{fig-lattice}. The two spin states experience different lattice potentials, which are given by
\begin{equation}
V_\uparrow(x) = V_\uparrow\cos^2(k_Lx) ~,~
V_\downarrow(x) = V_\downarrow\sin^2(k_Lx+\phi) ~.
\label{eq-trap}
\end{equation}
Here $k_L=\pi/a$, with $a$ being the lattice constant for each sublattice. One way to realize such a spin-dependent lattice potential is to employ two counter-propagating laser beams, which are linearly polarized with the two polarization vectors form an angle $\pi/2-\phi$ with respect to each other,
to form the so-called lin $\angle$ lin configuration \cite{sp-lattice-1}.
The spin-dependent AC stark shift gives rise to the potential described in Eq.~(\ref{eq-trap}). When $\phi=0$, the two sublattices are completely out of phase; whereas when $\phi=\pi/2$, the two sublattices are identical. As we will show, this angle $\phi$, which characterizes the spatial offset between the two sublattices, is an important control parameter, and, without loss of generality, we will restrict $|\phi| \in [0, \pi/4]$ in our discussion.

In such a lattice potential, two types of tunnelings are at present in general.
The first one is the \textit{intra}-sublattice tunneling, which originates from the atom kinetic energy.
The second is the \textit{inter}-sublattice tunneling, which can be induced by an additional radio-frequency (rf) field that drives a transition between the two spin states. This rf-induced inter-sublattice tunneling occurs between two adjacent sites, each from two different sublattices.
The Hamiltonian that describes our lattice model can be expressed in the following form,
\begin{align}
H=& \sum_j \Big[ \Delta \big(c_{j\uparrow}^\dag c_{j\uparrow} - c_{j\downarrow}^\dag c_{j\downarrow}\big)
-\Big( \sum_{\sigma=\uparrow\downarrow} t_\sigma c_{j\sigma}^\dag c_{j+1,\sigma}  \notag\\
&+J_1c_{j\downarrow}^\dag c_{j\uparrow}
+J_2c_{j\downarrow}^\dag c_{j+1,\uparrow} +\mathrm{H.c.} \Big) \Big]
 ~,
\label{eq-hamitonian_origin}
\end{align}
where $\{c_j,c_j^\dag\}$ are annihilation and creation operators of atoms on the $j$th lattice site,
$\Delta$ is the frequency detuning of the rf field from the bare transition frequency between the two spin states,
$t_\sigma$ is the intra-sublattice tunneling amplitude for each spin. To make our discussion more focused, we consider the situation where the two sublattices have equal amplitude, i.e., $V_\uparrow=V_\downarrow\equiv V_L$ in Eq.~(\ref{eq-trap}),
which leads to equal intra-sublattice tunneling amplitude with $t_\uparrow=t_\downarrow\equiv t$.
In Hamiltonian (\ref{eq-hamitonian_origin}), the inter-sublattice tunneling amplitudes along opposite directions are denoted as $J_1$ and $J_2$, whose relative magnitude can be controlled by the offset angle $\phi$. For $\phi=0$, we have $J_1=J_2$; for other values of $\phi$, they  differ from each other,
yielding a staggered inter-sublattice tunneling. More specifically, we have
\begin{equation}
\begin{split}
J_1=\Omega\int \mathrm{d}x \, W^*(x+a/2+\phi/k_L)W(x)\equiv J_1(\phi) ~,\\
J_2=\Omega\int \mathrm{d}x \, W^*(x-a/2+\phi/k_L)W(x)\equiv J_2(\phi) ~.
\end{split}
\label{eq-tunnel}
\end{equation}
Here $\Omega$ represents the rf field strength, and
$W(x)$ is the Wannier wave function of each sublattice.

The spin-dependent lattice model described by Hamiltonian (\ref{eq-hamitonian_origin}) underlies rich topological phenomena.
We shall show and discuss them in the following sections. 
The recoil energy is defined as $E_R=\hbar^2/2ma^2$, and will be chosen as the energy unit in the following.

\section{Topological features}\label{sec-topo}

\begin{figure}[htbp]
\centering
\includegraphics[width=0.48\textwidth]{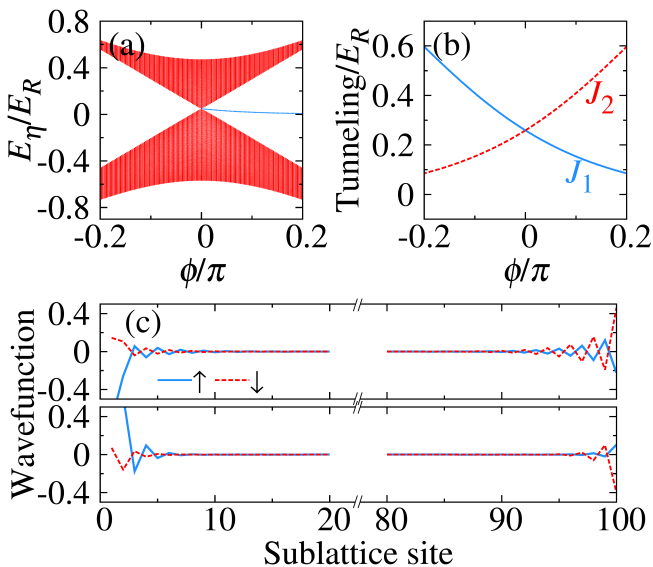}
\caption{(Color online)
(a) Energy spectrum of the generalized SSH model.
The blue solid line shows energy of the edge states.
(b) Inter-sublattice-tunneling amplitudes $J_1$ (blue solid line)
and $J_2$ (red dashed line) as functions of the phase $\phi$.
(c) The spatial distributions of two degenerate edge states at $\phi=0.2\pi$.
In our calculation, we set 100 sites for each sublattice.
The trap depth is set as $V_L=9.0E_R$
with the intra-sublattice-tunneling amplitude $t=0.0242E_R$.
The rf field strength $\Omega=2.0E_R$.}
\label{fig-ssh-topo}
\end{figure}

Hamiltonian (\ref{eq-hamitonian_origin}) features topologically nontrivial properties.
We first consider a case that the detuning $\Delta =0$.
Under this situation, our Hamiltonian describes the generalized Su-Schrieffer-Heeger (SSH) model \cite{ssh-chen}.
The SSH model was first proposed in the study of 1D polymer chains.
In its original form, the intra-sublattice tunneling $t$ is absent \cite{ssh-1,ssh-2}.
This can be easily achieved by a deep lattice trap.
In the past decades, the SSH model has attracted tremendous interest because of its rich topologically nontrivial features
such as topological solitons \cite{ssh-3}, and fractionally charged excitations \cite{ssh-4}. The SSH model has been successfully realized in a cold-atom system trapped in a superlattice potential where each lattice site takes the form of a double-well potential \cite{zak-phase,ssh-5}.
Different from it, our proposal provides an alternative simple route taking advantages of the sublattice spatial offset.
To show the topologically nontrivial properties of the lattice model,
we diagonalize Hamiltonian (\ref{eq-hamitonian_origin}) in the presence of nonzero $t$ into the form
\begin{equation}
\widetilde{H}=\sum_{\eta} \big( E_\eta \alpha_\eta^\dag \alpha_\eta -\frac{1}{2} \big) + \mathrm{const} \,.
\end{equation}
Here we perform the calculation with the open boundary condition by taking each sublattice to be of equal length with 100 sites.
We shall discuss in Sec.~V the case that the two sublattices possess different number of sites.
In Fig.~\ref{fig-ssh-topo}(a), we plot the energy spectrum for different $\phi$ values. As long as $\phi \neq 0$, hence $J_1 \neq J_2$ (see Fig.~\ref{fig-ssh-topo}(b)), there exists a two-fold degenerate zero-energy mode between the gapped bulk states. By examining the wave functions associated with the zero-energy modes, which is shown in Fig.~\ref{fig-ssh-topo}(c), we can identify them as the two topological edge states localized near the boundaries of the system.

To understand the origin of the topological features,
we transform Hamiltonian (\ref{eq-hamitonian_origin}) into the momentum space, 
\begin{equation}
H(k_x)=
\left(\begin{array}{cc}
-2t\cos(k_x a) & J_1+J_2 e^{-ik_xa} \\
J_1+J_2e^{ik_xa} & -2t\cos(k_x a)
\end{array}\right) ~,
\label{eq-ssh}
\end{equation}
where we have chosen the base $\Psi_{k_x}\equiv(c_{k_x\uparrow}, c_{k_x\downarrow})^T$.
We first make a unitary transformation $\Psi_{k_x}'=U\Psi_{k_x}$ with $U=\exp(i\hat{T}t)$,
to eliminate the kinetic energy term $\hat{T}\equiv-2t\cos(k_x a)$,
which does not change the topological properties of the system \cite{topo-cri-kinetic}.
Thus, in the rotating frame, the effective Hamiltonian becomes identical to that of the original SSH model
\begin{align}
H_\mathrm{SSH}(k_x) &= UH(k_x)U^{-1} - i U\partial_t U^{-1} \notag\\
&=
\left(\begin{array}{cc}
0 & J_1+J_2 e^{-ik_xa} \\
J_1+J_2e^{ik_xa} & 0
\end{array}\right) ~.
\label{eq-ssh2}
\end{align}
It is not difficult to see that the Hamiltonian (\ref{eq-ssh2}) respects the time-reversal symmetry
$\Theta H_\mathrm{SSH}(k_x) \Theta^{-1}=H_\mathrm{SSH}(-k_x)$ with $\Theta=\mathcal{K}$,
and the chiral symmetry $\Pi H_\mathrm{SSH}(k_x) \Pi =-H_\mathrm{SSH}(k_x)$ with $\Pi=\sigma_z$.
Here $\mathcal{K}$ is the complex conjugate operator and $\sigma_i~(i=x,y,z)$ are Pauli matrices.
Moreover, Hamiltonian (\ref{eq-ssh2}) preserves the inverse symmetry,
which implies $\sigma_x H_\mathrm{SSH}(k_x) \sigma_x =H_\mathrm{SSH}(-k_x)$.
Therefore, just like the SSH model, the topological features of the generalized SSH model can be described by a topological insulator belonging to the BDI class \cite{topo-cri-1,topo-cri-2,topo-cri-3}, even though Hamiltonian (\ref{eq-ssh}) itself does not obey the chiral symmetry.
The topological transition can be characterized by the Zak phase,
which is measurable in recent experiments \cite{zak-phase}:
\begin{equation}
\varphi_\mathrm{Zak} = i\int_{-k_L}^{k_L} \mathrm{d}k_x\, 
\langle \widetilde{\Psi}_{\pm}(k_x) | \partial_{k_x} | \widetilde{\Psi}_{\pm}(k_x) \rangle ~.
\end{equation}
Here $| \widetilde{\Psi}_{\pm}(k_x) \rangle$ are wave functions of the higher ($+$)
and lower ($-$) bands after diagonalizing Hamiltonian (\ref{eq-ssh2}).
We can easily obtain $\varphi_\mathrm{Zak}=\pi/2$
for $J_1>J_2$, yielding a gapped topologically trivial phase,
while $\varphi_\mathrm{Zak}=-\pi/2$ for $J_1<J_2$, yielding a topologically nontrivial phase
that hosts two degenerate edge states with a gapped bulk \cite{zak-phase}.
In our proposal, $\phi$ is a tunable parameter in experiments and controls the inter-sublattice tunneling $J_1$ and $J_2$,
see Eqs. (\ref{eq-tunnel}) and their amplitudes in Fig.~\ref{fig-ssh-topo} (b).
Tuning $\phi$ induces a topological phase transition at $\phi=0$.

We remark that,
in a more general case,
the intra-sublattice tunneling $t_\sigma$ can be made spin-dependent if $V_\uparrow\neq V_\downarrow$ in Eq.~(\ref{eq-trap}).
This will break the inverse symmetry, as discussed in Ref.~\cite{ssh-chen}.
However, the topological edge modes are robust against such a broken symmetry,
only resulting into two split branches.

\begin{figure}[htbp]
\centering
\includegraphics[width=0.48\textwidth]{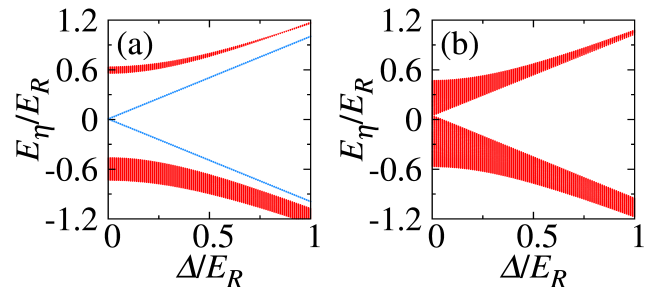}
\caption{(Color online)
Energy spectrum of the lattice model for (a) $\phi=\pi/5$ and (b) $\phi=0$.
We set 100 sites for each sublattice.
Other parameters are the same as in Fig.~\ref{fig-ssh-topo}, and the corresponding values of $J_{1,2}$ at a given $\phi$ can be inferred from Fig.~\ref{fig-ssh-topo}(b).}
\label{fig-rm-topo}
\end{figure}

\begin{figure}[htbp]
\centering
\includegraphics[width=0.45\textwidth]{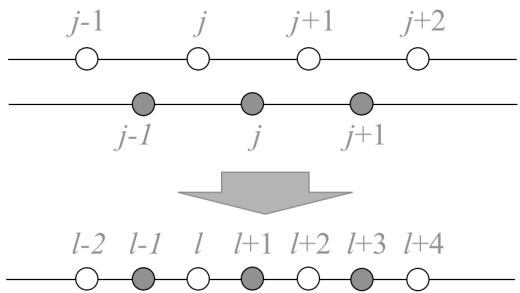}
\caption{Site index mapping from the 1D spin-1/2 model (top panel)
to the 1D spinless chain model (bottom panel).}
\label{fig-rm-model}
\end{figure}

Figure \ref{fig-rm-topo} illustrates the energy spectrum when the detuning $\Delta \neq 0$.
The system remains topological as long as $J_1<J_2$.
With the increase of $\Delta$,
the degeneracy of the edge modes is broken and split into two branches, as shown in Fig.~\ref{fig-rm-topo}(a).
In order to better understand the topological properties of the system,
we redefine the fermion operator index with the following mapping,
\begin{equation}
c_{j\uparrow} \rightarrow a_{2j}=a_l ~,\quad
c_{j\downarrow} \rightarrow a_{2j+1}=a_{l+1} ~, \label{eq-map}
\end{equation}
where $a_l$ represents the annihilation operator for a fictitious spinless fermion on the $l$th lattice site \cite{p-wave}.
In the new index representation,
the original 1D spin-1/2 lattice is mapped into a chain of spinless fermions.
The spin index $\uparrow$ or $\downarrow$ of the original lattice system corresponds to the even or odd site in the new model,
which is illustrated in Fig.~\ref{fig-rm-model}.
Hamiltonian (\ref{eq-hamitonian_origin}) is thus mapped into the following form,
\begin{align}
\mathcal{H}=& \sum_l \Delta(-1)^{l} a_l^\dag a_l - \sum_l\Big\{\big[ J+ \delta (-1)^l  \big]
a_l^\dag a_{l+1} \notag\\
&+t a_l^\dag a_{l+2} +\mathrm{H.c.}
\Big\} ~,
\label{eq-rm}
\end{align}
where detuning $\Delta$ corresponds to an effective staggered Zeeman field,
$J=(J_1+J_2)/2$, and $\delta=(J_1-J_2)/2$ characterize the staggered nearest-neighbor tunneling, and
$t$ describes the next-nearest-neighbor tunneling.
When $t=0$, Hamiltonian (\ref{eq-rm}) describes the Rice-Mele (RM) model \cite{rm-1,zak-phase,pump-1,pump-2}.
Similar to the SSH model, the RM describes a conducting polymers with nontrivial topology and has been realized in cold-atom systems confined in superlattice potential \cite{pump-1,pump-2}.
Its topological properties have been well studied in earlier works \cite{rm-2}. The presence of finite $t$ does not change qualitatively the topological features \cite{topo-cri-kinetic}.

\section{measurement of topological invariant} \label{sec-measure}

From Sec. \ref{sec-topo} we know that
the topological property of the 1D lattice system can be characterized by the Zak phase.
If the lattice offset angle $\phi$ is treated as a synthetic dimension,
the Zak phase in fact describes the Berry phase picked up
by a particle moving in closed trajectories across the first Brillouin zone \cite{zak}.
In our 1D lattice model,
the system remains unchanged under the transformation $\phi\rightarrow\phi+\pi$.
It indicates that in the synthetic dimension $\phi$, the system is periodic and should therefore respect the Bloch theorem.
Thus in the $k_x$-$\phi$ space,
the Berry phase for the $(\pm)$ band of Hamiltonian (\ref{eq-ssh}) can be given by \cite{berry}
\begin{equation}
\gamma_\pm = \frac{1}{\pi} \int_{\phi_0}^{\phi_0+\pi} \int_{-k_L}^{k_L} \Omega_\pm(k_x,\phi) \,\mathrm{d}k_x \mathrm{d}\phi ~,
\end{equation}
where the Berry curvature is expressed as
\begin{align}
\Omega_\pm(k_x,\phi) &= \mathrm{Im} \Big[
\frac{ \langle \widetilde{\Psi}_{\pm}|\partial_\phi H| \widetilde{\Psi}_{\mp}\rangle
\langle\widetilde{\Psi}_{\mp}|\partial_{k_x}H| \widetilde{\Psi}_{\pm}\rangle}{[E_\pm(k_x)-E_\mp(k_x)]^2}  \notag\\
& \quad- (\phi\leftrightarrow k_x) \Big] ~.
\end{align}
Here $E_{\pm}(k_x)$ is the dispersion of the $(\pm)$ band.
If we adiabatically change $\phi$ with a period $\pi$,
our system will thus behave like a Thouless quantum pump,
as has been detected in recent experiments \cite{pump-1,pump-2}.
We want to point out that, in these previous cold-atom experiments \cite{pump-1,pump-2}, the SSH and the RM models were realized using spinless atoms confined in a superlattice potential. As such, the hopping coefficients $J_1$, $J_2$, and $\Delta$ are not independently tunable. In our realization with the spin-dependent lattice, by contrast, the relative magnitude of $J_1$ and $J_2$ (or more specifically, their difference
$J_1-J_2$), and $\Delta$ are independently controllable.
It provides the possibility to study the Thouless pump evolving along various trajectories of the $(J_1$$-$$J_2)$-$\Delta$ plane, as we demonstrate below.

\begin{figure}[htbp]
\centering
\includegraphics[width=0.48\textwidth]{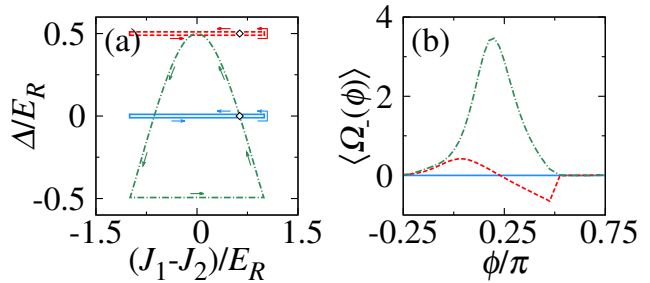}
\caption{(Color online) (a) Three paths of the quantum pump progress:
(i) blue solid line, $\Delta=0$; (ii) red dashed line, $\Delta=0.5E_R$;
(iii) green dash-dotted line, $\Delta=0.5\cos(2\phi)\,E_R$.
The dependence of $J_1-J_2$ on $\phi$ has been given in Eq. (\ref{eq-tunnel}).
The initial states of each pump trajectory are marked by the empty diamonds.
After $\phi$ evolves a $\pi$ period from $-\pi/4$ to $3\pi/4$, the system returns to each initial state.
The arrows describe the direction along each pump trajectory in the $(J_1$$-$$J_2)$-$\Delta$ plane.
(b) Evolution of the averaged Berry curvature $\langle \Omega_-(\phi)\rangle$ along the three trajectories in (a), respectively. The Berry curvature of the other branch is given by $\langle \Omega_+(\phi)\rangle=-\langle \Omega_-(\phi)\rangle$.}
\label{fig-pump}
\end{figure}

In Fig.~\ref{fig-pump} we show specific examples of the pump processes. The system is initially prepared in the state marked by the empty diamonds of Fig.~\ref{fig-pump}(a).
In order to detect the Berry phase of both ($\pm$) bands,
the atoms are initially loaded to occupy the two branches,
which can be prepared by controlling the atom number density of the system.
After varying $\phi$ by $\pi$, the system returns to the initial state so that a pump cycle is closed.
Here we consider three types of closed paths, shown in Fig.~\ref{fig-pump}(a):
(i) $\Delta$ remains at zero, i.e., the pump of the SSH model;
(ii) $\Delta$ is a nonzero constant that is independent from $\phi$, i.e., the pump of the RM model;
(iii) $\Delta$ evolves as $\Delta\sim\cos(2\phi)$.
In Fig.~\ref{fig-pump}(b), we plot the evolution of the averaged Berry curvature
$\langle \Omega_-(\phi)\rangle = \int\Omega_-(\phi,k_x) \mathrm{d}k_x$
along the three trajectories. The Berry curvature of the other branch is given by $\langle \Omega_+(\phi)\rangle=-\langle \Omega_-(\phi)\rangle$ \cite{berry}.
For the path (i), $\langle \Omega_-(\phi)\rangle$ vanishes during a pump period,
indicating that such a pump will not provide any information of the topological invariant of the system.
For the paths (ii) and (iii), $\langle \Omega_-(\phi)\rangle$ changes with $\phi$.
As the velocity of the atom current is determined by the Berry curvature $\Omega_\pm(k_x,\phi)$ \cite{berry, pump-1,pump-2,pump-3,pump-4,pump-5},
the center-of-mass position of the atom cloud will split into two branches and move along opposite directions.
The displacement between the two branches during a pump cycle is proportional to $\gamma_+-\gamma_-$.
For the path (ii), we find that $\gamma_\pm =0$.
Therefore, this pump progress cannot detect the topological invariant of the RM model.
In contrast, for the path (iii), the evolution of $\langle \Omega_-(\phi)\rangle$ leads to $\gamma_\pm = \mp\pi$.
As it has been discussed in Sec.~\ref{sec-topo},
$(J_1-J_2,\Delta)=(0,0)$ is the degeneracy point where the band gap closes, see Fig.~\ref{fig-ssh-topo}(a).
Only the trajectory that encircles this degeneracy point, i.e., the critical point of the topological transition,
can detect a quantized number \cite{pump-1},
which is the topological invariant of the SSH model.
Therefore, the pumping process, along any trajectory that encloses the degeneracy point, can facilitate the measurement of the topological invariant of the SSH model.

\section{Boundary effect}\label{sec-site}

\begin{figure}[htbp]
\centering
\includegraphics[width=0.48\textwidth]{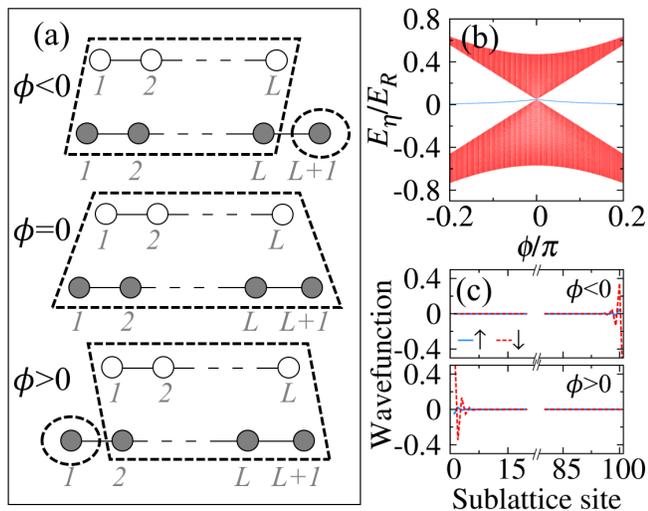}
\caption{(Color online) Effects of unequal numbers of sublattice sites on the topological properties.
We set $L=100$.
(a) Optical lattice configuration for different $\phi$: the empty or solid circles represent the spin $\uparrow$ or $\downarrow$ atoms.
(b) Energy spectrum of the system as a function of $\phi$. The system is topologically nontrivial with a single non-degenerate edge mode
as long as $\phi \neq 0$, i.e., $J_1 \neq J_2$.
(c) The spatial wave function of the edge mode at $\phi=\mp0.1\pi$ for the upper or lower panel. Here the edge mode is localized at opposite ends of the chain for the sign of $\phi$. Other parameters are the same as in Fig.~\ref{fig-ssh-topo}.}
\label{fig-boundary}
\end{figure}

In Sec. \ref{sec-topo},  we have shown that $J_1<J_2$ if the offset angle $\phi>0$, and $J_1>J_2$ if $\phi <0$, and the lattice system, described by the generalized SSH model, is topologically nontrivial in the former case, and trivial in the latter.
However, we note that we take the number of sites to be equal in both sublattices in our previous discussion.
Here we consider a situation where the number of sites are different for the two sublattices. Specifically, we take the number of sites for the spin-up sublattice to be $L$, and that for the spin-down sublattice to be $L+1$ as shown in Fig.~\ref{fig-boundary}(a). The energy spectrum as a function of $\phi$ is plotted in Fig.~\ref{fig-boundary}(b), from which we see that as long as $\phi \neq 0$ (i.e., $J_1 \neq J_2$), the system is topologically nontrivial with a finite bulk gap and a non-degenerate zero-energy edge mode.

That the edge mode is non-degenerate can be explained as follows.
Here we have an odd number of total lattice sites $2L+1$.
Following the discussion in Sec.~\ref{sec-topo},
we first make the unitary transformation with $U=\exp(i\hat{T}t)$ where $\hat{T}\equiv \sum_{j\sigma}\big(t_\sigma c_{j\sigma}^\dag c_{j+1,\sigma} + \mathrm{H.c.}\big)$,
to eliminate the $\hat{T}$ term in the Hamiltonian (\ref{eq-hamitonian_origin}), as this term does not change the topological properties of the system.
Thus, the generalized SSH Hamiltonian (\ref{eq-hamitonian_origin}) with $\Delta=0$, becomes that of the original SSH model,
\begin{equation}
H_\mathrm{SSH}= \sum_j \sum_{\sigma=\uparrow\downarrow}\big(J_1c_{j\downarrow}^\dag c_{j\uparrow}
+J_2c_{j\downarrow}^\dag c_{j+1,\uparrow} +\mathrm{H.c.} \big) ~,
\label{eq-hamitonian_ssh}
\end{equation}
whose form in the momentum space has been given in Eq.~(\ref{eq-ssh2}).
The system,
described by the Hamiltonian (\ref{eq-hamitonian_ssh}),
respects the chiral symmetry, implying that for each eigenstate $\alpha_\eta$ of the Hamiltonian (\ref{eq-hamitonian_ssh}) with energy $E_\eta$,
we can obtain another state $\hat{U}\alpha_\eta$ with the opposite energy $-E_\eta$ via a unitary transformation $\hat{U}$ \cite{topo-cri-3}.
Therefore, the spectrum of the Hamiltonian (\ref{eq-hamitonian_ssh}) hosts $L$ sets of energy pairs $\{ E_\eta,-E_\eta\}$,
leaving a single zero-energy mode $E_{L+1}=0$ as the edge mode.
For the original Hamiltonian, which includes the $\hat{T}$ term, the above analyses are still valid, as $\hat{T}$
only results in an energy shift to the energy spectrum, see Fig.~\ref{fig-boundary}(b).
Therefore, the single zero-energy mode is robust unless the chiral symmetry of the Hamiltonian (\ref{eq-hamitonian_ssh}) is broken.
Further examination shows that this edge mode is localized at the right (left) end of the chain if $\phi<0$ ($\phi>0$), as shown in Fig.~\ref{fig-boundary}(c). As $|\phi| \longrightarrow 0$, the edge mode becomes less localized as its spatial width increases. At $\phi=0$, the bulk gap closes, the edge mode dissolves into the bulk, and the system becomes topologically trivial. Therefore, in this situation, $\phi=0$ remains as a critical point for the topological phase transition. However, unlike the situation depicted in Fig.~\ref{fig-ssh-topo}, here the system is topologically nontrivial at both sides of this critical point, with the corresponding edge state localized at opposite ends of the chain.

\section{Artificial magnetic field}\label{sec-flux}
\begin{figure}[htbp]
\centering
\includegraphics[width=0.48\textwidth]{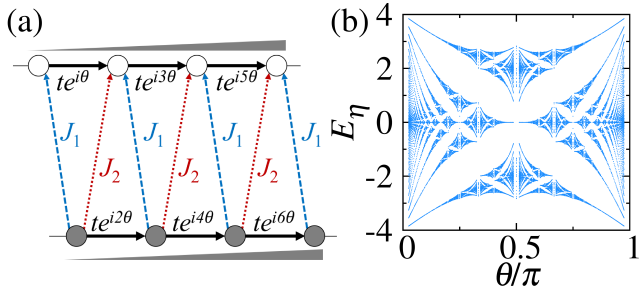}
\caption{(Color online) (a) Optical lattice configuration:
Raman-assisted tunneling in each linear tilted sublattice with an energy offset
hosts a spatial dependent phase.
Gray triangles along the chain denote the tilted directions.
(b) Hofstadter butterfly: energy spectrum $E_\eta$ versus the particular flux $\theta$.
$E_\eta$ is in units of $t$ and we set $J=t$.}
\label{fig-flux}
\end{figure}

Recent experiments have demonstrated the creation of artificial magnetic field in optical lattice systems using the technique of laser-assisted tunneling \cite{laser-hop-1,laser-hop-2}. This technique can be implemented in our system, and the resulting model describes an effective 2D triangular lattice. 
To see this, we first suppress the conventional intra-sublattice tunneling by introducing a tilt along the lattice direction (for simplicity, we assume the tilt is spin-independent, hence the two sublattices are tilted in the same way.). This can be achieved, for example, by including gravity along the same direction, or by adding a magnetic gradient field.
Subsequently, we restore the intra-sublattice tunneling by two laser beams with different wave number $k_1$ and $k_2$ along the lattice direction, and
with a frequency difference $\omega_1-\omega_2$ matching the energy different between adjacent lattice sites with the same sublattice.
Here we consider a simple setup with offset angle $\phi=0$.
After time averaging over the rapidly oscillating terms,
we obtain a complex intra-sublattice tunneling amplitude: $t \exp\big(i\cdot (2j-1)\theta\big)$ for spin $\uparrow$
and $t \exp\big(i\cdot 2j\theta\big)$ for spin $\downarrow$
with $\theta\equiv (k_1-k_2) \cdot a/2$, as shown in Fig. \ref{fig-flux} (a).
The inter-sublattice tunneling is still induced by the rf field.

For this situation, if we exploit the fermion operator index mapping (\ref{eq-map}) again,
the Hamiltonian is expressed as 
\begin{equation}
H=-\sum_l \big( J a_{l}^\dag a_{l+1} + t e^{il\theta} a_{l}^\dag a_{l+2} \big) ~,
\label{eq-hamit-flux}
\end{equation}
where we assume $J_1=J_2=J$.
From Hamiltonian (\ref{eq-hamit-flux}), the lattice system indeed mimics a magnetic field
with a tunable flux $\theta$ around each triangular plaquette as shown in Fig.~\ref{fig-flux} (a).
Different from the conventional magnetic field, the flux in each triangular cell induced by the synthetic magnetic field is staggered, rather than homogeneous \cite{flux1}.
The corresponding energy spectrum as a function of $\theta$ 
is plotted in Fig.~\ref{fig-flux} (b).
We can clearly find a fractal pattern if $\theta=\pi p/q$ ($p$, $q$ are prime integers), which is known as the Hofstadter butterfly
as a feature of the synthetic magnetic field \cite{flux2}.
If periodic boundary condition is implemented,
the energy band for each flux $\theta=\pi p/q$ is split into $q$ sub-bands.
Therefore, our lattice model provides an alternative route to synthesize the magnetic fields in a 1D system,
and can be utilized in studying various intriguing effects \cite{flux1,flux2,flux3,flux4,flux5}.

We note that similar artificial magnetic field can be realized in our model by keeping the intra-sublattice tunneling conventional, but induce the inter-sublattice tunneling with a pair of Raman beams instead of an rf field. In this way, $J_1$ and $J_2$ become complex.

\section{Discussion}\label{sec-summary}

In summary, we have proposed a protocol to realize a wide variety of topological models, based on a spin-dependent optical lattice potential. We have shown how unconventional tunneling, staggered Zeeman field, and artificial staggered magnetic flux can be readily engineered with our platform.
In this realization, many of the key parameters can be independently controlled, and various boundary conditions can be constructed. Therefore this realization allows us to explore a variety of exotic lattice models that are difficult to be realized in conventional condensed matter systems.

In the present work, we have focused on the single-particle physics.
If we introduce the $s$-wave interaction between the two spin states, we can have effective $p$-wave interaction as was demonstrated in our earlier work \cite{p-wave}. For nonzero offset angle, the strength of this effective interaction becomes staggered, which may give rise to unconventional phases. The interaction effects, combined with the single-particle manipulation covered here, represent a very interesting line of future research.

\section{Acknowledgements}

Z.Z., X.Z., and G.G. are supported by National Natural Science Foundation of China (Grants No. 11474271 and No. 11674305).
H.P. acknowledges support from the U.S. NSF and the Welch Foundation (Grant No. C-1669).
Z.Z. thanks Chunlei Qu and Bin Wang for helpful discussions,
and acknowledges support from National Postdoctoral Program for Innovative Talents of China (Grant No. BX201600147).

\end{document}